\documentstyle[epsfig,here,12pt]{article}

\makeatletter
\@addtoreset{equation}{section}
\makeatother

\begin{document} 
\newcommand {\ber}
{\begin{eqnarray*}}
 \newcommand {\eer} {\end{eqnarray*}}
\newcommand {\bea}
{\begin{eqnarray}} \newcommand {\eea} {\end{eqnarray}} \newcommand {\beq}
{\begin{equation}} \newcommand {\eeq} {\end{equation}}
\newcommand {\eqref} [1] {(\ref {#1})}

\def\II{\relax{\rm 1\kern-.22em I}}

\def\Acknowledgements{\bigskip  \bigskip {\begin{center} 
             \bf ACKNOWLEDGEMENTS \end{center}}}

\def\cS#1{\footnote{S:#1}}
\def\ti{{\tilde \imath}}

\begin{titlepage}
\titlepage
\rightline{hep-th/0102007}
\rightline{CPHT-S001.0101}
\vskip 3cm
\centerline{{\Large \bf On non-commutative ${\cal N}=2$ super Yang-Mills}}
\vskip 1.5cm
\centerline{\bf Adi Armoni$^a$, Ruben Minasian$^a$ and Stefan Theisen$^b$}
\centerline{armoni,ruben@cpht.polytechnique.fr, theisen@aei-potsdam.mpg.de}
\begin{center}
\em $^a$Centre de Physique Th{\'e}orique, {\'E}cole 
Polytechnique\footnote{Unit\'{e} mixte du CNRS et de l'EP, UMR 7644}
\\91128 Palaiseau Cedex, France
\end{center}
\begin{center}
\em $^b$Max-Planck-Institut f{\"u}r Gravitationsphysik, Albert-Einstein-Institut,
\\Am M{\"u}hlenberg 1,D-14476 Golm, Germany
\end{center}
\vskip 1.5cm
\begin{abstract}
We discuss the Seiberg-Witten solution of the non-commutative
${\cal N}=2$ $U(N)$ SYM model. The solution is described in terms of
the ordinary Seiberg-Witten curve of the $SU(N)$ theory plus
an additional free $U(1)$. Hence, at the two-derivative
approximation the theory flows to the ordinary commutative theory in
the infra-red ($k<1/\sqrt{\theta}$).  
In particular, the center $U(1)$ is free and it
decouples from the other $U(1)'s$. In addition, no UV/IR mixing
is found.  
\end{abstract}

\vfill
\begin{flushleft}
{\today}\\
\end{flushleft}
\end{titlepage}

\newpage

\section{Introduction}
Non-commutative field theories have recently attracted much attention,
mainly due to the discovery of their connection to string theory
\cite{Connes:1998cr,Seiberg:1999vs}.
The perturbative structure of these theories is interesting
\cite{Minwalla:1999px}. 
Though they differ from ordinary theories
by higher derivative terms, perturbative results seem to indicate   
that they do not flow to the ordinary theory in the infrared. 
Contributions from the UV affect the IR. This surprising behavior,
called UV/IR mixing, seems to contradict the Wilsonian picture.
We usually refer to higher dimensional operators as 'irrelevant' in
the IR. However, in the case of non-commutative theories, an infinite
sum of irrelevant contributions conspire to be relevant in the IR.

The picture just described rests on one-loop calculations 
\cite{Matusis:2000jf}.
In order to gain a complete understanding of the IR, 
non-perturbative methods might be required. In the case of ${\cal N}=4$
SYM, it was found, by using the AdS/CFT correspondence, that there is no
UV/IR mixing and the theory converges to the commutative one in the IR 
\cite{Hashimoto:1999ut,Maldacena:1999mh}.
The reason is that the non-planar
contributions, which are responsible to the UV/IR mixing, cancel in this
case \cite{Matusis:2000jf}.

A model for analyzing the infra-red behavior of the
non-commutative theories is pure ${\cal N}=2$ SYM. This theory is not
finite and \`a priori we expect UV/IR mixing. Moreover, using
the Seiberg-Witten theory\cite{Seiberg:1994rs}
we can analyze the IR physics. Several
attempts in this directions were already made. In the first
\cite{Sheikh-Jabbari:2000en} the fate of the center of the gauge group
was not addressed. We disagree with the results of the two other
papers \cite{Yoshida:2000xt,Bellisai:2000aw}. 

The one-loop effective action of the NC $U(1)$ ${\cal N}=2$ SYM was
 recently derived \cite{Khoze:2000sy,Zanon:2000nq}.
 Based on this one-loop calculation, it was suggested
 \cite{Khoze:2000sy} that the $U(1)$ theory flows asymptotically to a
 free theory in the IR. Note, however, that non-perturbative effects
 might be essential for a generic value of $\theta$. \footnote{ 
Shortly after the first version of our paper was published in hep-th, 
the paper \cite{Hollowood:2001ng} appeared. 
The authors arrive at the same conclusion that 
we have put forward, namely that the $SU(N)$ part is unchanged compared 
to the commutative case and that the $U(1)$ part is free in the IR
at least at the two derivative level to which we have restricted our 
analysis.}

In order to describe the theory at the IR, we parametrize it as an
{\em ordinary} theory, namely in terms of a holomorphic prepotential.
We assume higher derivatives to be irrelevant.
The non-commutativity and the UV/IR mixing should
enter via the prepotential. Note that an IR description of the $U(N)$
theory in
terms of a set of non-commutative $U(1)'s$ does not make sense since
the non-commutative $U(1)$ theory is asymptotically free.
Thus, we assume an ordinary (commutative) description in the IR.
 
The summary of our results is the following:
we find that the two-derivative effective action flows
in the IR to the commutative one and there is no UV/IR mixing. 
Namely, the $U(N)$ theory is 
described in terms of a free $U(1)$ (the center) and a set of
$N-1$ additional $U(1)'s$ which are described by the ordinary
Seiberg-Witten curve. In particular, we do not find in the IR any
coupling between the center $U(1)$ and the other $U(1)'s$. We should
emphasize, however, that this is probably due to the two-derivatives
approximation that we use. At higher derivatives we do expect a
coupling of the center $U(1)$ and the rest of the $U(1)'s$.

The organization of the manuscript is as follows: section 2 is devoted
to field theory analysis whereas section 3 is devoted to string theory
analysis. In section 2.1 we review
the perturbative behavior of the model. In section 2.2 we describe
the IR behavior of the $U(1)$ theory. Section 2.3 is devoted to a
discussion about the expected behavior of the $U(N)$ model. In
section 2.4 we present its solution. Section 3.1 and 3.2 are devoted to 
the description of the solution from the type IIA and M-theory points
of view. It is used to support our findings in field theory.  
  
\section{Field Theory Analysis} 

\subsection{Perturbative Dynamics}

Non-commutative gauge theories exhibit an interesting perturbative
behavior. First, it is important to note that, within the framework
of \cite{Seiberg:1999vs} only certain gauge
groups and representations 
can be consistently described over the non-commutative space.
In particular we cannot discuss $SU(N)$ theories, only $U(N)$. 
The reason is that the $SU(N)$ theory is not gauge invariant under the
generalized non-commutative gauge transformation since it contains 
anti-commutators of generators \cite{Madore:2000en,Matsubara:2000gr};
see however also \cite{Wessetal}.

The planar graphs of the non-commutative theories are exactly the 
same as the planar graphs of ordinary theory, apart from overall
global phases \cite{Filk:1996dm}.
On the other hand, non-planar graphs are believed to be
UV finite \cite{Bigatti:2000iz}.
However, these graphs lead to a new kind of infra-red
divergences \cite{Minwalla:1999px}.
Since the source of these new type of divergences is at
high momenta, there is an interesting UV/IR mixing. For supersymmetric
gauge theories the only new IR divergences are logarithmic, in
contrast to the quadratic IR divergences in the non-supersymmetric
theory \cite{Matusis:2000jf}.
At one loop, these effects appear only in amplitudes which involve
external legs in the center of the gauge group \cite{Armoni:2001xr}.
{}For ${\cal N}=4$
theory theses effects cancel \cite{Matusis:2000jf} and in the 
infrared the theory converges 
to the ordinary one \cite{Hashimoto:1999ut,Maldacena:1999mh}.
 
It is important to
note that even though the non-commutative field theories have been shown 
to have a `nicer' UV behavior than the 
corresponding commutative theories, a proof of renormalizability
has not been given yet.
However, since these theories can be realized as a vacuum of
string theory via brane configurations, they should be consistent at
the quantum level, even in the field theory limit. (The simplest way
of realizing non-commutative ${\cal N}=2$ theories
 in type IIA, is to consider type IIB with D3 branes on a
$Z_2$ orbifold singularity with a constant NS-NS two-form background and
then T-dualize to have D4 branes suspended between NS5 branes
compactified on a circle \cite{Douglas:1996sw}).

We assume that the ${\cal N}=2$ model is renormalizable (one loop
evidence for the renormalizability of the model is given
in \cite{Aref'eva:2000bg}
). If the model is renormalizable, the
structure of divergences should be the same as of the planar
commutative theory. Namely, the same counter-terms are needed to
regularize the theory and the beta-function consists of only the one
loop contribution which is \cite{Armoni:2001xr}
\beq
\beta = N_f - 2N
\eeq
for an $U(N)$ theory with $N_f$ hypermultiplets. In particular the
pure $U(1)$ theory is asymptotically
free \cite{Martin:1999aq,Krajewski:2000ja}.
We will restrict ourselves to
theories without matter, though the generalization to the cases 
when matter is present is straightforward.
 
\subsection{The ${\cal N}=2$ $U(1)$ Non-Commutative SYM}

${\cal N}=2$ non-commutative Yang-Mills can be formulated in
superspace similarly to the ordinary theory \cite{Ferrara:2000mm} (see also 
\cite{Terashima:2000xq} and \cite{Zanon:2000gy,Santambrogio:2000rs}
for a recent application to the ${\cal N}=4$ case).
The  ${\cal N}=2$ gauge multiplet contains a complex
scalar field in the adjoint representation. 
In contrast to the commutative theory, the NC theory is not free and
there is a potential for the scalar. It is
\beq
V\propto (\phi \star \bar \phi - \bar \phi \star \phi)^2,
\eeq
where the square is with respect to the star product. 
As usual, the classical moduli space of vacua is determined by 
finding physically inequivalent static solutions of $V=0$.

Assuming that the minima of the 
potential are unchanged by quantum corrections,
they are
\beq
\phi = a  ={\rm constant}.
\eeq
There is, however, a crucial difference between the commutative and the
non-commutative theories. 
Usually, in the ordinary $SU(N)$ theory there is a moduli space
parameterized by $vev$'s of 
gauge invariant monomials of the scalar fields. 
Here, giving a $vev$ breaks the gauge symmetry spontaneously.  
In the $U(1)$ non-commutative case the transformation
\beq
\phi \rightarrow \phi + a, \label{shift}
\eeq
with constant $a$ is a symmetry of the theory, i.e. there is no moduli 
space which is parameterized by $a$. 
Therefore, not only
the vacuum is insensitive to this shift, but all the physics.
For example, there are no $W$-boson masses which depend on $a$. 
Note that this symmetry holds also at the quantum level. This happens since
both the action and the measure are invariant under the shift (as we shall see
in the type IIA picture this corresponds to the fact that we have the
freedom to change the location of the D4 branes without changing the
physics on these branes). 

One may therefore wonder what the theory in the IR is.

The symmetry \eqref{shift} is spontaneously broken in the vacuum. The
realization
of this breaking is the appearance of a Goldstone boson. We assume
here that the Goldstone theorem holds in the non-commutative case 
\cite{Petriello:2001mp}. One
might wonder if this is indeed true, as the proof of Goldstone theorem
assumes locality. However, non-commutativity is a mild non-locality, and in
particular, we observe that there is a Goldstone boson in the UV
regime. It is the scalar itself. The realization of the gauge theory
via type IIA would give us further support that a Goldstone mode exists.
By completion of the ${\cal N}=2$ multiplet, we arrive to the conclusion
that we should have a massless ${\cal N}=2$ vector multiplet in the
IR, as well. What is therefore the form of the low-energy affective action ?

To lowest order in derivatives, namely to quadratic order, the action 
can be written in terms of a prepotential ${\cal F}$
\beq
\int d^2 \theta {\partial ^2 {\cal F} (A) \over \partial A ^2}
W_\alpha W^\alpha . \label{U1action}
\eeq

Since the shift \eqref{shift} is a symmetry of the theory, the
effective
coupling cannot depend on $a$. It means that the prepotential must
be of the following form
\beq 
{\cal F} = {1\over 2}\tau _0 A^2 ,
\eeq
and at the quadratic level the IR theory is free, exactly as the
commutative one. At first sight this looks surprising, since the UV theory
is asymptotically free. However, note that the action \eqref{U1action}
describes the dependence on the 'moduli' $A$ at the IR and not the
coupling constant as a function of the energy. 
Namely, though the gauge coupling runs, we cannot read its running
from the prepotential. The reason is that the $vev$ of the scalar is
{\em not} related to the energy scales, in contrast to the ordinary
case where $W$ bosons exist.

We may recall that the standard way of deriving the one-loop prepotential 
is to find an effective action 
which reproduces the $U(1)_{\cal R}$ anomaly
(see e.g. \cite{Alvarez-Gaume:1997mv}).
Supersymmetry requires the divergence of the $U(1)_{\cal R}$
to be controlled by the one-loop beta function. 
In the present case the one-loop beta function 
does not vanish and thus neither does the ${\cal R}$-anomaly
(see \cite{Ardalan:2000qk} for a related
discussion). Neither of them is smooth in the $\theta \rightarrow
0$ limit. A quadratic
prepotential, however, cannot reproduce the $U(1)_{\cal R}$ anomaly.
Nevertheless, we claim that there is no contradiction
between the shift symmetry and the existence of the $U(1)_{\cal R}$ anomaly.
Moreover, a naively expected $\log A$ term in the effective
action would, as in \cite{Seiberg:1999vs}, 
require additional singularities which signal
the appearance of massless states. We find such a scenario
unlikely.

\subsection{The $U(N)$ case}
Now we turn to the case of non-commutative $U(N)$ Yang-Mills theory.
At the generic point on the moduli space the theory is broken to 
$U(1)^N$. The infra-red theory is described by $N$ independent 
vector multiplets. In order to see this, let us use the following
generators to describe the Cartan subalgebra
\beq
{\tilde T} ^\ti = {\rm diag} (0,..,0,1,0,...,0)
\label{tt}
\eeq
Accordingly there are $N$ photons ${\tilde A} ^\ti _\mu$, 
$\ti=1,\dots,N$.  
It is important
 to observe that the residual non-commutative gauge symmetry
 does not mix the photons
${\tilde A} ^i _\mu$. Indeed, under NC gauge transformation, each photon
transform into itself in the following way
\beq
{\tilde A }^\ti _\mu  \rightarrow {\tilde A}^\ti _\mu  + \partial _\mu
\lambda ^\ti + \sin (\theta^{ \nu \rho} \partial ^{(1)} _\nu \partial
^{(2)}
 _\rho )
{\tilde A}^\ti _\mu \lambda ^\ti ,
\eeq
where the derivative $\partial ^{(1)}$ acts on the photon and the
derivative $\partial ^ {(2)}$ acts on the gauge parameter.
The non-commutative parameter has only non-vanishing space-space
components, e.g. $\theta^{12}\neq0$.

The IR theory thus consists of $N$ photons. However, the low-energy
action should be parameterized by only $N-1$ $vev$'s. The reason is
that the $vev$ of the 'center of mass' $U(1)$ is not a modulus.
This is as in sect. 2.2. In order to see
this, it is better to parameterize the Cartan by the traceless generators
\beq
T^i = {\tilde T}^\ti - {\tilde T}^{\ti+1} = {\rm diag} (0,...,0,1,-1,0,...0)
\label{t}
\eeq
and by the 'center of mass' generator $T^0= \II$.
The theory is invariant under a shift of the 'center of mass'
scalar by a constant.

Thus the low energy effective action should be written in the
following way
\beq 
\int d^2\theta\left({\partial^2{\cal F}(A,A_{\{i\}})\over\partial A ^2}
W_\alpha W^\alpha+\sum_{ij}{\partial^2{\cal F}(A,A_{\{ i\}})\over\partial
A_i\partial A_j }
W_\alpha^i W^{\alpha^j}\right ) ,
\label{UNaction}
\eeq
where $A$ stands for $A_0$. By the same arguments as in the previous
section it is clear that the prepotential should be of the 
form
\beq
{\cal F} = {1\over 2}\tau _0 A^2 + f(A_{\{ i \} }) \label{pre}
\eeq
Therefore, the center of mass $U(1)$ is decoupled from the rest of 
 the $U(1)'s$. This is precisely the same behavior as in the 
 commutative case. The $U(N)$ commutative theory decouples to a
trivial (center) $U(1)$ and $U(1)'s$ which are associated with the broken
$SU(N)$. The difference is that the non-commutative theory in the UV
does not split into two non-interacting 
pieces. Rather, the non-commutative $U(1)$ interacts
with the $SU(N)$ part. Only at low energies such a decoupling occurs.

\subsection{The Seiberg-Witten curve}
With the understanding \eqref{pre} we can proceed in a similar way as in
the commutative case.
The generalization of the Seiberg-Witten curve to the $U(N)$ case
seems to be \cite{Klemm:1995qs,Argyres:1995xh,Martinec:1996by}
\bea
& &
y^2 -2 P(z)y + 1 =0 \\
& &
P(z)=\prod_{\ti=1}^N(z-a_\ti)=z^N-\sum_{i=0}^{N-1} u_i z^{N-i} \label{pol}
\eea
where, in particular, $u_0=\sum a_i$.
In the type IIA picture discussed below, the D4 branes end on the NS5
branes at the positions $a_i$ and $u_0$ is proportional to their 
center-of-mass position. 

The kinetic term if the effective action can be written as \cite{deBoer:1998zy}
$$
S=\int d^4 x K_{i\bar\jmath}\partial_{\mu}u_i \partial^{\mu}\bar
u_{\bar\jmath}
$$
with $i,j=1,\dots,N$. Where
\beq
K_{i\bar\jmath} = \int _{\Sigma} d^2 z {z^{N-i-1} \over y - P(z)}
{\overline {z^{N-j-1} \over y - P(z)}}
\eeq

Note that $K_{0\bar 0}$, i.e. the kinetic energy associated 
with the center-of-mass, diverges. In fact, by going to center-of-mass
coordinates via the shift $z\to z+{1\over N}\sum a_i$ we can eliminate
$u_0$ so that \eqref{pol} effectively only depends on $N-1$ parameters. 
Thus, the solution at low-energy reduces to the ordinary
solution, at the two-derivatives approximation. 
An indication
that this is indeed the case is given by the gravity solution
\cite{Buchel:2000cn}. Note however that the gravity solution is 
valid only at large $N$ and therefore the issue of the 'center of mass' 
$U(1)$ cannot be addressed in this approach.

At first look it seems to contradict the UV/IR mixing of the
non-commutative theories. There are one-loop non-planar graphs,
associated with the $U(1)$, which indicate a mixing of the UV/IR.
In fact, the one-loop effective action analysis which takes into account 
the UV/IR mixing also suggests a flow to a free $U(1)$ theory \cite{Khoze:2000sy}. 
 We suggest that at very low energies, the
${\cal N}=2$ theory flows to a commutative theory.

\section{The picture from string theory}

\subsection{Type IIA}
The realization of non-commutative ${\cal N}=2$ SYM is as 
in the commutative case, but with an additional constant NS-NS two-form
along the $1,2$ components. The brane configuration consists of
$N$ coincident $D4$ branes in the $0,1,2,3,6$ directions and two
parallel NS5
branes in the $0,1,2,3,4,5$. The D4 branes spans a finite segment of
the 6-direction and the NS5 branes are located at the ends of this
segment \cite{Witten:1997sc}.

The type IIA pictures captures the classical theory and the one-loop
effects \cite{Witten:1997sc}. It does not describe the IR theory, only the UV. Let us see how it describes the $U(N)$ non-commutative theory.

First of all, it describes really $U(N)$ and not $SU(N)$. The reason
is the following. Let us assume that the D-branes are separated. In
the presence of NS-NS field there is a modified potential. The
potential, due to their positions, is
\beq
{\rm tr} \ 
(\phi  \star \bar \phi - \bar \phi \star \phi )^2\,.
\eeq
Let us shift the position of the branes by a 4d space-time
dependent piece. 
\beq
\phi \rightarrow \phi + a(x).
\eeq
Without the NS-NS background (or, equivalently, 
without the star product), this shift
does not affect the potential, only the kinetic term of the $U(1)$.
However, the modified potential is not invariant under this shift. It
means that the $U(1)$ field is not decoupled. On the other hand, when
$a$ is constant, it is still a symmetry. One can always shift the
D-brane positions by a constant. That means that the $vev$ is not a
modulus. This shows that the symmetry \eqref{shift} holds at the
quantum level, at least perturbatively. 
The fixing of the position of the branes breaks this symmetry and the
 result is a Goldstone boson, exactly the same as in the ordinary case.

The bending of the NS5 brane is not affected by the NS-NS 4d
background. It describes the dependence of the gauge coupling on the 
$vev$ of the scalars. This is why the one-loop beta function of the
non-commutative theory is the same as the commutative one.
The case of a single D4 brane (the $U(1)$ theory) is exceptional. In
this case, even in the commutative theory, the NS5 branes bend.
However, it does not imply a running of the gauge coupling as a
function of the energy since a $vev$ (a separation of D4 branes) is
needed to relate the bending to the ``real'' beta function.

\subsection{M-theory}

In the M-theory picture the brane configuration becomes a single
M5. The configuration is non-compact and the non-compactness carries
the additional $U(1)$, which is located at 'infinity'.
Thus the low energy theory (even in the
commutative case) is in fact $U(1)^N$ and not $U(1)^{N-1}$.
The point is that the additional $U(1)$ is not dynamical. No $W$-bosons
are associated with it. Therefore, the conclusion in the commutative
case is that this additional $U(1)$ must be free.

In order to realize non-commutativity, a constant 3-form background should
be added \cite{Bergshoeff:2000jn}.
However, the 3-form background does not change the
geometrical picture. The $U(1)$ case is described by a single
M5 with zero genus. The $U(1)$ mode that lives on the zero genus M5
is non-normalizable and hence it
 supports our conclusion that the $U(1)$ theory becomes free in the infrared. 
Let us turn now to the $U(N)$ theory. 
The compact part of the 
M5 is still a genus $g=N-1$ Riemann surface. So, it seems
that also in this background there are only $N-1$ dynamical photons
and the additional $U(1)$ decouples from the dynamics. 

We now provide some details.  First we briefly summarize the structure of
the theory (see for details \cite{Klemm:1996bj,Witten:1997sc}). The
harmonic decomposition of the self-dual two-form on
the M5 worldvolume gives rise to $U(1)^g$ gauge fields, while two out of
five scalars, which parameterize the position of the curve $\Sigma$ in the
four-dimensional manifold $Q={\bf R}^3 \times S^1$ 
(${\bf R}^3=\lbrace x^4,x^5,x^6\rbrace$)  
should be decomposed in the
basis of the deformations of the normal bundle thus giving $g$ complex
scalars needed to complete ${\cal N}=2$ vector multiplets. Finally, an
extra
scalar arises from the "decomposition"  of the two-form in terms of the
volume form of the finite part of $\Sigma$ (alternatively seen as a
Hodge-dual of the two-form in four dimensional space-time.).
 This compact mode joins
the three remaining scalars on the fivebrane worldvolume to form a
"universal" hypermultiplet. As is \cite{Minasian:2000qn}, one
can show explicitly by looking at the
reduction of susy transformations that this multiplets decouples from the
rest of the theory. 

 Dealing with the non-compactness of $\Sigma$ as in \cite{Witten:1997sc},
one expands the field-strength of the two-form in terms of the 
harmonic one-forms $\omega_i$ on $\Sigma$ as
\beq
H = \sum_{i}^g \left( F_i \wedge \omega_i + *F_i \wedge *\omega_i \right)
\label{hd}
\eeq
to find at the linearized level that the dynamics is governed by the
intersection matrix $\tau_{ij}=\int_\Sigma \omega_i\wedge *\omega_j$. 
One has to bear in mind that in the
full theory $H$ is not closed and is related to a closed three-form in a
complicated non-linear fashion. While this nonlinearity greatly
complicates the full analysis, it is not very important in the far
infrared regime where we are working and where the theory is broken to
$U(1)^g$. Most importantly, turning on the bulk $C$-field does not lead to
additional deformations of this sector, and it is precisely this
deformations that we are interested in. As we will argue now the only
deformation induced by the $C$-field affects the center of mass $U(1)$,
which also unfortunately is the weakest point of the M-theory derivation
of SW theory (See e.g. \cite{Giveon:1999sr}). 

The one-form, $\omega_0$,  related to the
center of mass $U(1)$, requires special attention. This is the only
one-form on
$\Sigma$ that extends to $Q$ (note that $\omega_0$ is fixed on
$\Sigma$) and hence is orthogonal to $\omega_i$
($\tau_{0i}=0$). Moreover when extended to the bulk it should coincide
with the 
only one-form on $Q$, namely the form supported on the M-theory circle.
In order to obtain the relevant coupling, with a slight abuse of notation
and omitting the pull-back signs,  we take $C=B \wedge \omega_0$. The
$C$-field coupling to M5, $\int_{M_6}
\vert H-C \vert^2$, gives a coupling to the center of mass $U(1)$
while leaving the other $U(1)$ fields unaffected:
\beq
\int_{M_4} \vert F_0-B \vert^2.  
\label{coup}
\eeq 
As mentioned before the coefficient in front of this term is infinite but
in the far infrared we can absorb it by a field redefinition
\footnote{ As already mentioned we are not discussing the matter fields,
but at this point it is natural to ask what happens in situations
corresponding to type IIA limit where we have
multiple fivebranes. While the M-theoretic discussion, in particular
the assertion that there is only one $U(1)$ field which can be coupled to 
the NS two-form via \eqref{coup}, does not
change, at first sight the situation looks much different both from IIA
and field theory point of view. However it is not hard to see that even in
the case of multiple fivebranes (product gauge groups) there is only {\it
one} Goldstone field and thus only one additional massless $U(1)$. Indeed,
mutual motion of the groups of fourbranes in the adjacent areas of the
interior of the fivebrane chain affects the masses of the hypermultiplets
\cite{Witten:1997sc},
and thus the only allowed shift now is that of the whole system. 
Field-theoretically speaking, one can also see that any shift, other than
simultaneous shift in all scalars, will change the mass of the
bifundamental hypers due to the change in the Yukawa couplings.}.
Unfortunately this argument only shows that the $C$-field coupling deforms
only the center of mass $U(1)$ and that in the far infrared it does not
affect the remaining gauge fields, confirming our field-theoretical
findings. This conclusion is in complete agreement with the
supergravity analysis \cite{Buchel:2000cn} which shows that
 the commutative and 
the non-commutative moduli spaces coincide at large $N$. 
However since the adequate description
of the center of mass is notoriously difficult in this picture, it does
not shed much light on the nature of the $U(1)$ theory itself, and here we
will have to rely solely on field theory arguments.

\Acknowledgements
We would like to thank  L. Alvarez-Gaume, C. Bachas, M. Douglas,
F. Hassan, Y. Oz and A Schwimmer for useful discussions. 
Interesting conversations with
O. Aharony, J. Barbon, M. Berkooz, 
S. Elitzur, K. Lansteiner, E. Lopez, N. Nekrasov, E.
Rabinovici, S. Shatashvili and S. Yankielowicz are also gratefully acknowledged.
The work of A.A. and R.M. is supported in part by EEC contract
HPRN-CT-2000-00122.
The work of S.T. is supported by GIF - the German-Israeli Foundation 
for Scientific Research and by European Commission RTN programme 
HPRN-CT-2000-00131 in which he is associated to U-Bonn.

\end{document}